\documentstyle[11pt,newpasp,twoside,epsf]{article}
\reversemarginpar

\begin{document}
\title{Theory of Formation of Massive Stars via Accretion}
 \author{Harold W. Yorke}
\affil{Jet Propulsion Laboratory, California Institute of Technology,\\
 MS 169-506, 4800 Oak Grove Drive, Pasadena, CA 91109}

\begin{abstract}
The collapse of massive molecular clumps can produce high mass
stars, but the evolution is not simply a scaled-up version of
low mass star formation. Outflows and
radiative effects strongly hinder the formation of massive stars
via accretion.  A necessary condition for accretion growth
of a hydrostatic object up to high masses $M \ga 20\,$M$_\odot$
(rather than coalescence of optically thick objects) is the
formation of and accretion through a circumstellar disk. Once
the central object has accreted approximately 10 M$_\odot$
it has already evolved to core hydrogen-burning; the resultant
main sequence star continues to accrete material as it begins
to photoevaporate its circumstellar disk (and any nearby disks)
on a timescale of $\sim$10$^5\,$yr,
similar to the accretion timescale. Until the disk(s) is (are)
completely photoevaporated, this configuration is observable as
an ultra-compact H{\sc ii} region (UCH{\sc ii}).  
The final mass of the central star (and any 
nearby neighboring systems) is determined by the interplay
between radiation acceleration, UV photoevaporation, stellar
winds and outflows, and the accretion through the disk. 

Several aspects of this evolutionary sequence have been
simulated numerically, resulting in a "proof of concept".
This scenario places strong
constraints on the accretion rate necessary to produce high
mass stars and offers an opportunity to test the accretion
hypothesis.
\end{abstract}

\section{Introduction}
Massive stars play a key role in the sequence of events after
the Big Bang that ultimately result in the development of life
on Earth.  They are the principal source of heavy elements and of
UV radiation and are an important source of dust grains.  Through
a combination of winds, massive outflows, champagne flows, and
supernova explosions they provide an important source of turbulence
in the ISM of galaxies.  This turbulence in combination with
differential rotation presumably drives galactic dynamos.  The
galactic magnetic fields thus produced, interacting with supernova
shock fronts, accelerate cosmic rays.  Cosmic rays, UV radiation,
and dissipation of turbulence are the principal sources of heating
in the ISM, whereas heavy elements found in dust, molecules, and
in atomic/ionic form ultimately are responsible for its cooling.
Massive stars thus profoundly affect the star and planet
formation process as well as the physical structure of galaxies.

In spite of the key role that these relatively short-lived massive
stars play in the shaping of galactic structure and evolution,
our understanding of their formation is still rather limited.
The reason for this is three-fold:  They are difficult to observe
during the critical formation phases, they are rare, and the
theoretical problem is extremely complex.  Their formation is
obviously not merely a scaled-up version of low mass star
formation and obviously much more complex because of the
proximity of high mass stars (among these the forming star
itself), resulting in mutual interactions via gravitational
torques, powerful winds and ionizing radiation.

\begin{figure}
\plotone{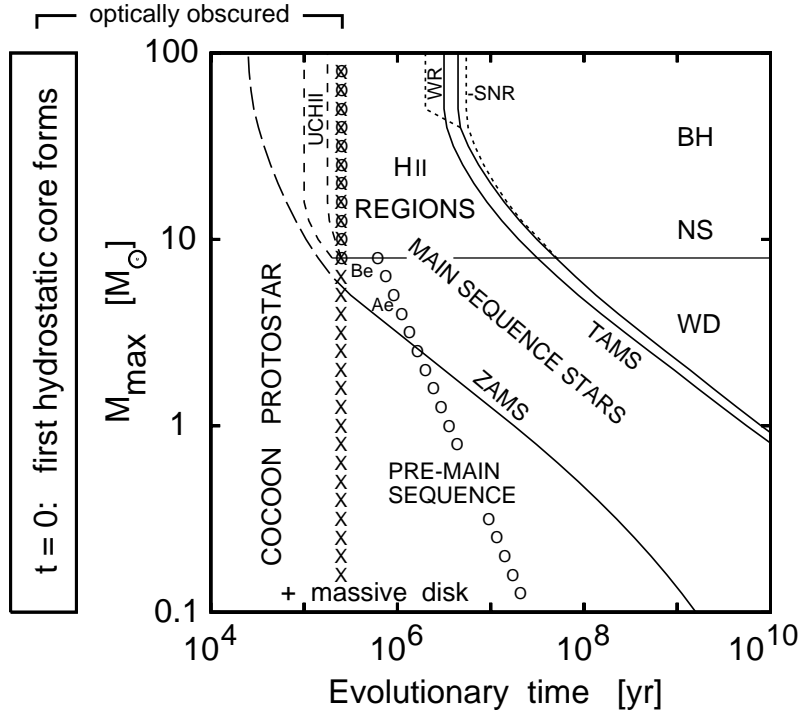}
\caption{Relevant time scales of stars attaining a maximum mass
$M_{\rm max}$.  Symbols have the following meaning: WR: Wolf-Rayet star,
TAMS: end of core hydrogen-burning, ZAMS: begin of hydrogen-burning,
Ae \& Be: Herbig AeBe stars; circles: dissipation of 
gaseous disks (my guess); crosses: end of rapid accretion onto
accretion disk (assumed independent of mass).
The end of nuclear burning (unmarked line parallel to the TAMS) 
results in a WD (white dwarf) or -- for massive stars -- in an SNR
(observable supernova remnant) and either a BH (black hole) or 
NS (neutron star) [adapted from Yorke 1986].}
\end{figure}

Not only are massive stars formed less often than their low 
mass counterparts,  their relevant time scales (contraction to 
hydrogen-burning and nuclear burning timescales) are shorter
(see Fig.~1).
Both effects result in fewer examples of high mass stars to be
found in a given (early) evolutionary phase within a given volume.
The low number statistics of high mass stars is partially
offset by their higher luminosities, which allow us to
study high mass stars at greater distances.  However,
insufficient spatial resolution is an issue --- an entire
cluster is often contained in a single observing pixel.


In the following I will not argue for
or against the ``accretion scenario'' versus the ``coalescence
scenario'' for high mass star formation,
but rather investigate conditions under which accretion of
material onto a massive star via an accretion disk is
theoretically possible.  The basic questions to be addressed are:
\begin{enumerate}
\item How is it possible to compress sufficient material to
      form a massive star into a sufficiently small volume
      within a sufficiently small time period?
\item How does forming massive star influence its immediate
      surroundings, eventually limiting the final mass?
\end{enumerate}
As McKee \& Tam (2003) have argued, turbulent and pressurized
clouds permit sufficient material to be available in the cores of
giant molecular clouds  for high mass star formation.
Here, we address the issue
of further concentrating this material into a region of a few
R$_\odot$ within a few 10$^5$~yr.

\section{The basic problem}
 
A necessary condition to accrete sufficient material to produce
a massive star ($M \ga 10$ M$_\odot$) is:
$
 M_* = \int_0^t \left[ \dot M_{\rm acc}(t') - \dot M_{\rm out}(t') 
 \right]\, dt' \ga 10\;{\rm M}_\odot
$,
i.e., the infall (accretion) rate $\dot M_{\rm acc}$ must greatly
exceed the outflow rate $\dot M_{\rm out}$ during a significant
proportion of the formation process.  For this
to occur the acceleration due to gravity must exceed
the outward directed radiative acceleration of the embryo source.
Whereas gravity $GM_*/r^2$
increases linearly with mass, the radiative acceleration of dusty
material $\kappa L/4\pi r^2 c$ is proportional to the
stellar luminosity which increases as a high power of stellar mass
(roughly $L_* \propto M_*^{3.2}$ in the range
$0.01 \la M_*/{\rm M}_\odot \la 100$).
Thus, to allow infall we require
$\kappa_{\rm eff} L / 4 \pi r^2 c < GM_* / r^2 $
with $ L = L_* + L_{\rm acc}$,
which translates into
\begin{equation}
  \kappa_{\rm eff} < 130 \; {\rm cm^2 \, g^{-1}} \;
  \left[ { M_* \over 10\; {\rm M}_\odot } \right]
\left[ { L \over 1000\; {\rm L}_\odot } \right]^{-1}
\end{equation}
This condition defines the maximum effective opacity $\kappa_{\rm eff}$
of accretable material. The (proto-)star's luminosity is given by the
sum of its intrinsic luminosity $L_*$ and the luminosity $L_{\rm acc}$
emitted by the dissipation of kinetic energy of the material being accreted.

Dusty material generally has a very high opacity (Fig.~2);
for the ``hardness'' of radiation expected from main sequence stars of
5 M$_\odot$ and higher, the net force on
typical dusty interstellar material ($\kappa \sim 100$~cm$^2$g$^{-1}$)
is directed away from the star (Fig.~3).

\begin{figure}
\plotone{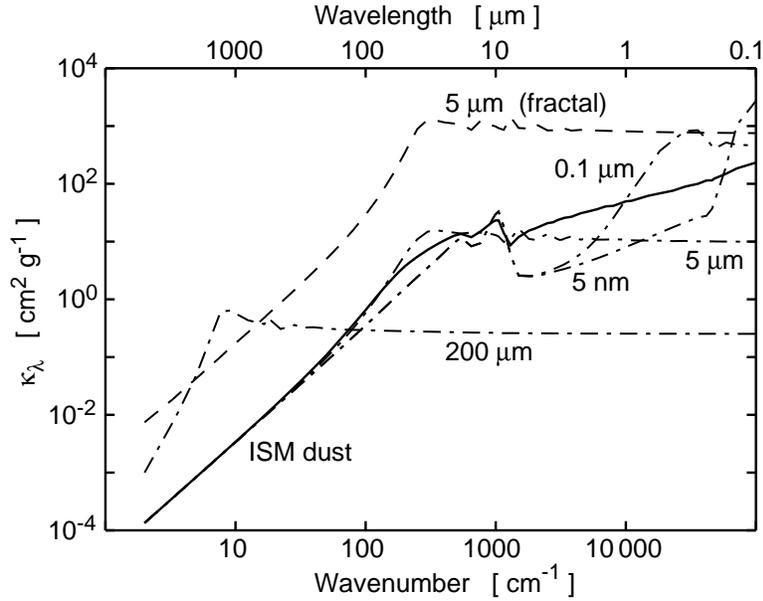}
\caption{Specific extinction coefficients of dusty gas
with a dust to gas mass ratio of 0.01
under the assumption of compact silicate grains of given radius
({\sl dashed-dotted lines}).
For comparison the opacity of fractal 5~$\mu$m silicates
({\sl dashed lines})
and ``ISM Dust'' ({\sl solid line}; Preibisch et al. 1993) are
also given.}
\end{figure}

\begin{figure}
\plotone{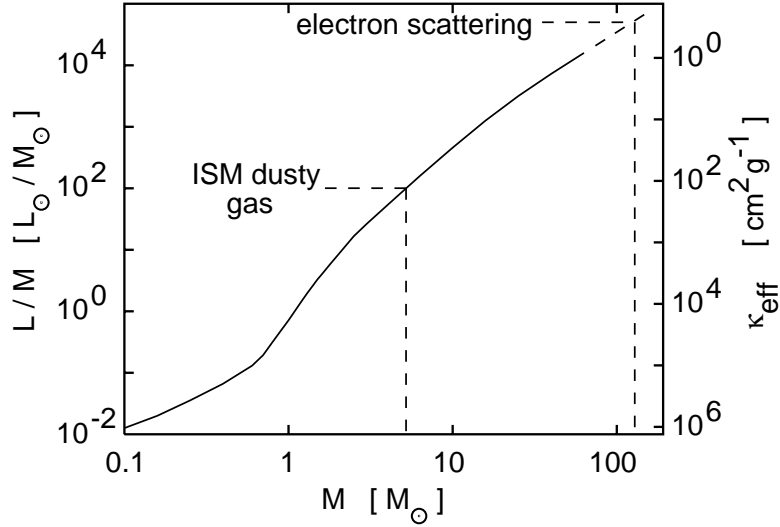}
\caption{Luminosity to mass ratio ({\sl solid line, left axis})
for main sequence stars.  Using equation~2 this translates
into a critical effective opacity for main sequence stars that allow
accretion of material ({\sl right axis}).  The dotted lines
show how two extreme values for opacity translate into upper 
mass limits of 5 M$_\odot$ (ISM dusty gas) and 130 M$_\odot$ 
(electron scattering).}
\end{figure}

\section{How does Nature solve this Problem?}

To allow further growth of an already existing stellar embryo at least
one of the following conditions must be met: a) $\kappa_{\rm eff}$
must be significantly lower than its ISM value for optical/UV radiation;
b) The effective luminosity must be reduced; or c) ``gravity'' must
be increased.
\\

\noindent
{\bf a) Reduce $\kappa_{\rm eff}$:} \\
As evident in Fig.~2 $\kappa_{\rm eff}$ can be significantly lower
than its ISM value if the radiation field ``seen'' by the accreting
material is shifted from the optical/UV into the far infrared, if the
average size of dust grains increases (but remains ``compact'' rather
than becoming ``fractal'') or if
most of the dust is destroyed.  In their pioneering efforts Kahn (1975)
and Wolfire \& Cassinelli (1987) studied the 1D spherically symmetric
accretion problem for massive star formation
with emphasis on the dust opacity. Indeed, the latter authors
concluded that massive stars can only form if the dust has been
significantly modified, assuming an accretion flow that is steady-state
and spherically symmetric.
Of course, accretion may be non-steady and/or non-spherically symmetric
and this basic premise may be invalid.

Another possibility to reduce the effective opacity is the accretion
of optically thick ``blobs''.  In this case
\begin{equation}
  \kappa_{\rm eff} = {\pi R^2_{\rm blob} / M_{\rm blob}}
\end{equation}
As a particular subset of this family of solutions
Bonnell et al. (1998) considered building
up massive stars by coalescence of lower mass stars within a stellar
cluster (see also Bonnell 2002).  

Modifications to the opacity due to 
coagulation of dust and dust destruction processes during the collapse
phase were calculated by Suttner \& Yorke (2001) for three different
detailed dust models (compact spherical particles, fractal BPCA grains,
and fractal BCCA grains).
Using a 2D (axial symmetry assumed) code that followed the
dynamics of gas + 30 individual dust components, they find that even during
the early collapse and the first $\sim$10$^4$~yr of dynamical disk
evolution, the initial dust size distribution is strongly modified
(Fig.~4). Close to the disk's midplane coagulation
produces dust particles of sizes of several 10\,$\mu$m (for compact spherical
grains) up to several mm (for fluffy BCCA grains), whereas in the vicinity
of the accretion shock front (located several density scale heights above
the disk), large velocity differences inhibit
coagulation. Dust particles larger than about 1\,$\mu$m segregate from
the smaller grains behind the accretion shock. Due to the combined effects of
coagulation and grain segregation the infrared
dust emission is modified. Within the accretion disk a
MRN (Mathis, Rumpl, \& Nordsieck 1977) dust
distribution provides a poor description of the general dust properties.
Nevertheless, the radiative force acting on the {\sl infalling}
material is hardly affected.

\begin{figure}
\plottwo{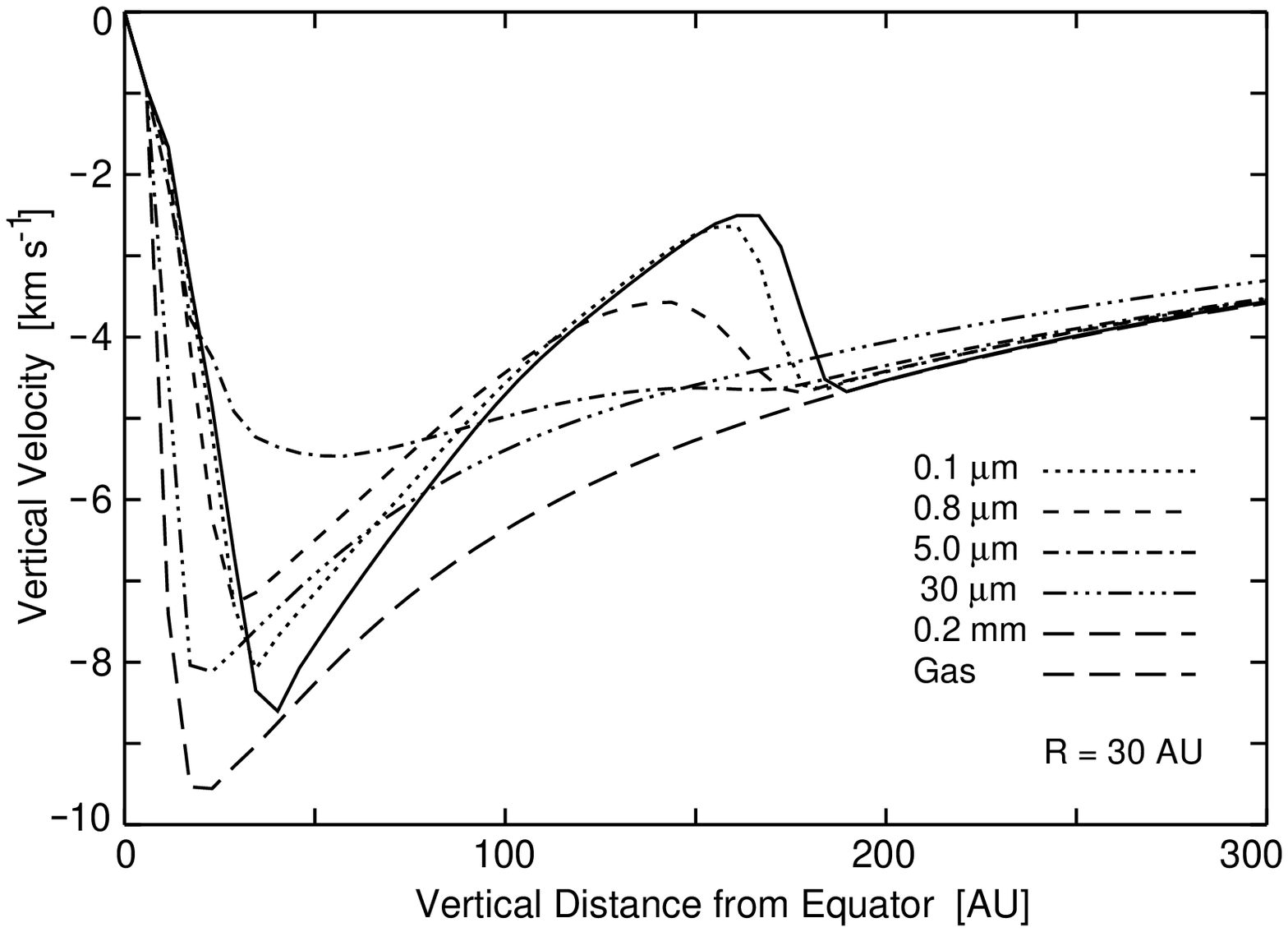}{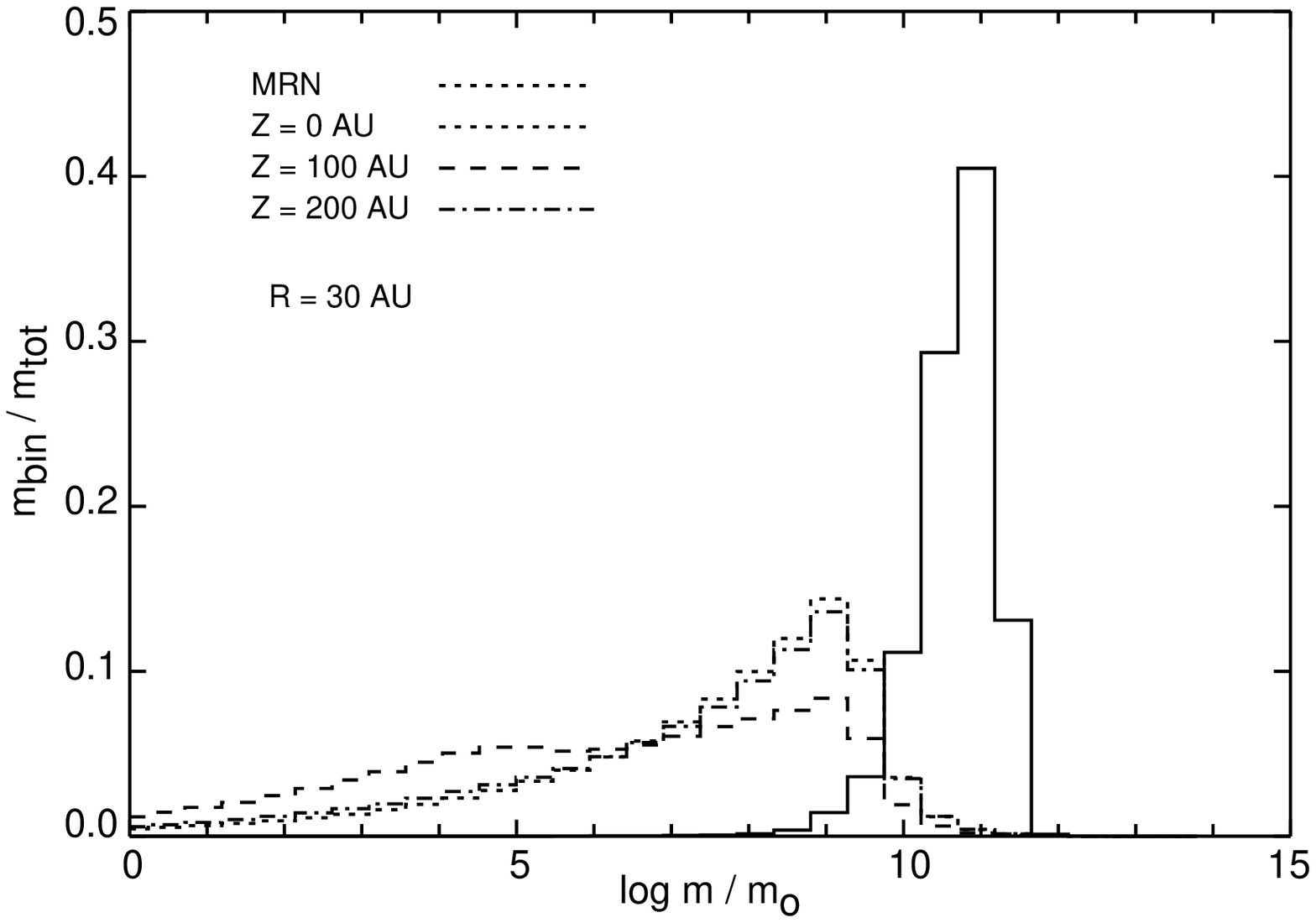}
\caption{Evolution of compact spherical grains in a rotating, collapsing
3~M$_\odot$ protostellar clump at 11,400~yr (Suttner \& Yorke 2001).
{\bf Left:} Velocities of selected grains through the accretion shock
at $r=30$~AU;
{\bf Right:} Grain mass spectrum at selected positions along $r=30$~AU
compared to initial MRN spectrum.}
\end{figure}

\noindent
{\bf b) Reduce the effective luminosity:} \\
Yorke \& Kr\"ugel (1977) solved the time dependent non-steady state
accretion problem in spherical symmetry and were able to produce stars
of masses 17 M$_\odot$ and 36 M$_\odot$ from clouds of masses 50 M$_\odot$
and 150 M$_\odot$ respectively --- due to the effects of oscillatory
``super-Eddington'' accretion.  Accretion was permitted during
quiescent low luminosity phases.  Also, the sheer weight of the
entire dusty envelope forced material upon the star, even when the
Eddington criterion was not fulfilled locally.

Nakano et al. (1995), however, point out that even a small amount
of rotation leads to the formation of a circumstellar disk, and we thus
expect accretion to proceed in 2D through an accretion disk, i.e.
radiation pressure could blow away the tenuous polar regions
but not the massive disk.  Yorke \& Bodenheimer (1999) studied
this effect quantitatively.
They find that whereas the central object may emit radiation
isotropically, the radiation field quickly becomes anisotropic further
from the center. For an outside observer and in particular for a dust
grain attempting to accrete onto an existing protostellar disk, the 
radiative flux close to the equatorial plane can be much smaller than
the component parallel to the rotation.  This so-called ``flashlight
effect'' (the ``beaming'' of radiation in the polar direction) occurs
whenever a circumstellar disk forms.

As an example of the flashlight effect Fig.~5 displays the angle-dependent
SEDs of a 2~M$_\odot$ protostellar clump at two evolutionary times.
The edge-on bolometric fluxes are a factor of 0.2 and 0.07 (respectively)
less than what they would be for an isotropic source, whereas the pole-on
bolometric fluxes are a factor of 2.6 and 2.9 greater.  Moreover, the
edge-on flux is dominated by the far infrared, which is
far less effective at radiatively accelerating dusty gas than the mid- and
near-infrared seen pole-on.

\begin{figure}
\plotone{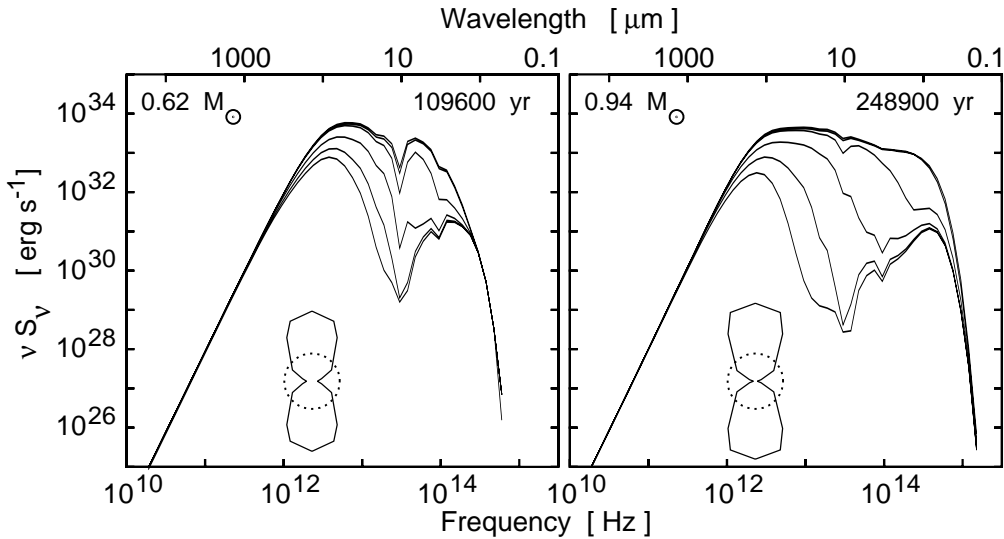}
\caption{SEDs of a 0.13~pc protostellar clump at viewing angles:
0$\deg$ (pole-on), 20$\deg$, 30$\deg$, 60$\deg$, 75$\deg$, and 90$\deg$.
Accreted core mass ({\sl upper left}) and
time after hydrostatic core formation ({\sl upper right})
are given in each frame. In insets
the ``beam pattern'' of the bolometric luminosity (vertical 
corresponds to pole-on) is compared
to that of an isotropic source ({\sl dotted line}) [adapted from 
Yorke \& Bodenheimer 1999].}
\end{figure}

Although the flashlight effect allows dusty material to come close
to the central source via a circumstellar disk, the
material to be accreted eventually encounters optical and
UV radiation from the central source.  A necessary requirement for
this material to be accreted rather than ``blown out'' by radiation
is that the dust has been largely destroyed (or it has coagulated
into larger particles) so that the opacity is dominated by the
gaseous component.

Even though no massive disk has yet been directly
observed around a main sequence massive star, there is much indirect
evidence that such disks exist (Shepherd, Claussen, \& Kurtz 2002).
In their radio recombination maser studies and CO
measurements Martin-Pintado et al. (1994) do find indirect
evidence for both an ionized stellar wind and a neutral disk
around MWC349. Moreover, several other high luminosity FIR sources
--- suspected embedded young OB stars ---
have powerful bipolar outflows associated with them (e.g., Eiroa
et al. 1994; Shepherd et al. 2002).  Such massive outflows
are probably powered by disk accretion, and, similar to their
low mass counterparts, the flow
energetics appear to scale with the luminosity of the source
(see Cabrit \& Bertout 1992; Shepherd \& Churchwell 1996; Richer
et al. 2000). 

If the primary source of the massive
star's material is from the surrounding molecular clump via accretion,
then a circumstellar disk should be the
natural consequence of the star formation process even in the high
mass case.  However, it should be difficult to observe disks around
massive stars.  The high FUV and EUV fluxes associated with high mass stars
will begin to photoevaporate the disks on timescales of $\sim$10$^5\,$yr
(Hollenbach et al. 2000), which will be observable as deeply embedded
UCH{\sc ii}s with comparable lifetimes (Richling \& Yorke 1997).
\\

\noindent
{\bf c) Increase Gravity:}
For completeness I mention the fact that the gravitational acceleration
is enhanced with respect to radiative acceleration when massive stars
form within a dense cluster of not so brightly radiating objects.  For
this to have a dominant effect we require that
$
  \rho_{\rm objects} \gg \rho_{\rm gas}
$.
In this scenario one requires a density-peaked
cluster of low mass objects embedded within a molecular cloud. The
effective gravity near the cluster's center is enhanced relative to
an isolated molecular cloud without the cluster and relative to 
off-center regions of the molecular cloud.  If this were the only way
to form massive stars, isolated massive stars may not
exist at all or only in very exceptional cases.

\section{What happens to the central (Proto-)Star during Accretion?}

Because the luminosity is so critical during accretion up to high
stellar masses, one must also consider the luminosity evolution of
the accreting object.  As discussed by Maeder (2002) and Yorke 
(2002) an initially low mass object that gains mass through accretion
evolves substantially differently in the Hertzsprung-Russell (HR) diagram
than a non-accreting protostar would (see Fig.~6). 
 
Yorke's (2002) tracks are qualitatively similar to the
more detailed calculations by Behrend \& Maeder (2001)
and Meynet \& Maeder (2000).  Differences can be attributed 
to the starting mass, 0.1~M$_\odot$ instead of 1~M$_\odot$, and the
differing accretion rates.  In all cases published to date,
not only do the tracks of accreting objects consistently lie slightly
below the equilibrium deuterium burning ``birthline'', but
the qualitative effect of more rapid accretion is to shift the
tracks to even smaller radii away from the ``birthline''.
Indeed, the concept of ``birthline'' is no longer valid for stars
more massive than $M_* \ga 0.7$~M$_\odot$, because deuterium is
consumed faster than it can be accreted.
The tracks of accreting stars eventually converge to the main sequence
and follow along the ZAMS as more material is added.  For e.g.
an accretion rate of 10$^{-3}$~M$_\odot$, hydrogen-burning begins at 
$t \simeq 1.3 \times 10^4$~yr, after $\sim$13~M$_\odot$ have accreted.

\begin{figure}
\plotone{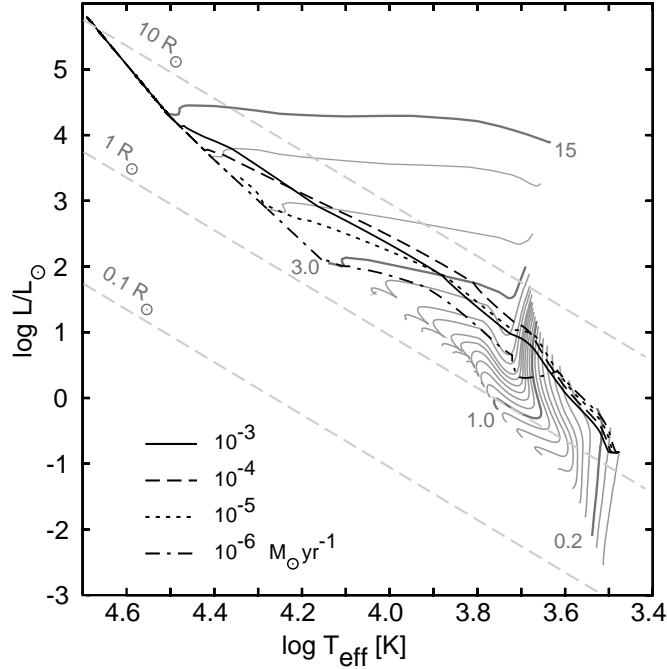}
\caption{Pre-main sequence tracks of accreting (proto-)stars in the
HR diagram for the given constant accretion rates are compared to published
tracks ({\sl grey lines}) of non-accreting pre-main sequence stars
(D'Antona \& Mazzitelli 1994; Iben 1965).
All ``accreting'' tracks are assumed to begin at the 
``birthline'' of an equilibrium deuterium burning 0.1~M$_\odot$
pre-main sequence star [adapted from Yorke 2002].}
\end{figure}

I remind the reader that these tracks in the HR diagram do not
reflect the actual observable bolometric luminosities of accreting
protostars.  Much of the accretion luminosity $L_{\rm acc}$
will be indistinguishable from the intrinsic luminosity $L_*$
of the star.  The importance of
accretion luminosity, $L_{\rm acc} \sim GM\dot M/r$ is exemplified
in Fig. 7. At $\dot M_{\rm cr} = L_* r / GM$ accretion luminosity
and intrinsic luminosity are equal ({\sl dash-dotted line} for
core hydrogen-burning stars and {\sl dash-triple dotted line} for birthline
stars). Above this line the luminosity is dominated by accretion;
below the line it is dominated by the intrinsic stellar luminosity.

What order of magnitude of mass accretion rate can be expected?
In order to produce a star of mass $M$ within, say
200,000~yr, an average accretion rate
$5 \times 10^{-5}$~M$_\odot$~yr$^{-1}$~$[M/{\rm M}_\odot]$ is 
necessary ({\sl dashed line} in Fig.~7).  Assuming this average accretion
rate during the main accretion phase, we note that the luminosity
of low mass stars is dominated by accretion luminosity,
whereas for high mass stars the luminosity is initially determined
by accretion but is eventually dominated by the intrinsic stellar
luminosity.  Of course the actual accretion rate may vary strongly
from this average (see example given in Fig.~8).
The maximum accretion rate possible onto a core hydrogen-burning star,
assuming electron scattering and the effects of both the intrinsic
stellar luminosity and accretion luminosity is given by the
{\sl solid line}.  This maximum allowable accretion rate is modified
for low mass stars if ``birthline'' stellar radii and luminosities
are assumed ({\sl dotted line:}).

\begin{figure}
\plotone{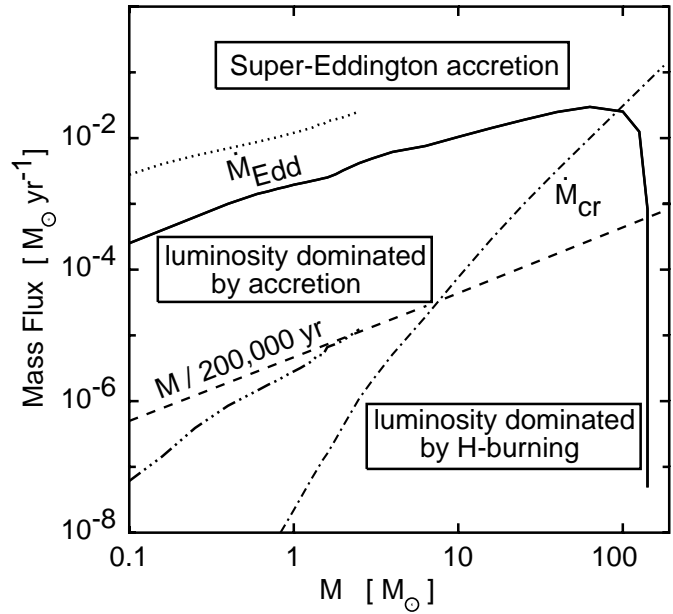}
\caption{Relevant mass accretion rates onto existing stars of mass $M$
as discussed in text.  }
\end{figure}

Yorke \& Sonnhalter (2002) consider the collapse of isolated,
rotating, non-magnetic, massive molecular clumps of masses 30~M$_\odot$
60~M$_\odot$, and 120~M$_\odot$ using an improved frequency-dependent
radiation hydrodynamics code (see Fig.~8).
The flashlight effect discussed in section 3
allows material to enter into the central regions through a disk.
For massive stars it is important to take into account the frequency
dependent nature of the opacity and the flux within the disk (rather
than assuming either Rosseland or Planck ``grey'' opacities).
For their 60~M$_\odot$ case Yorke \& Sonnhalter
find that 33.6~M$_\odot$ are accreted in the central regions
as opposed to 20.7~M$_\odot$ in a comparison
``grey'' calculation.  Because these simulations cannot spatially
resolve the innermost regions of the molecular clump, however, they
cannot distinguish between the formation of a dense central cluster,
a multiple system,
or a single massive object.  They also cannot exclude significant
mass loss from the central object(s) which may interact with the inflow
into the central grid cell.  With the basic assumption that all
material in the innermost grid cell accretes onto a single object,
they are only able to provide an upper limit to the mass of stars which
could possibly be formed.
 
\begin{figure}
\plottwo{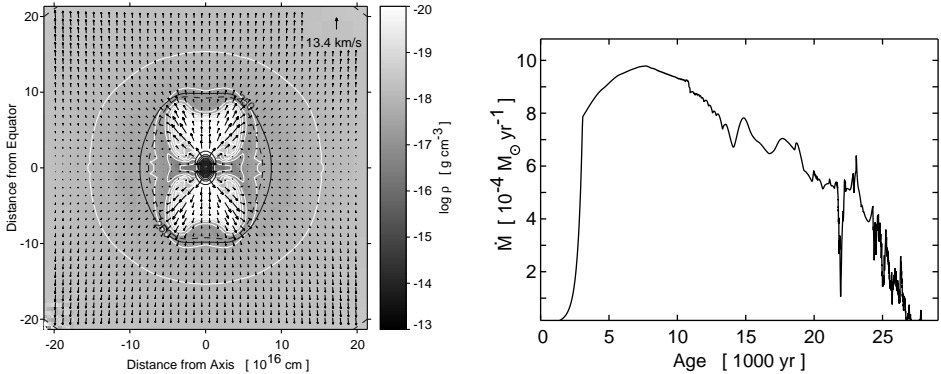} {fig8b.ps}
\caption{{\bf Left:} Distribution of density ({\it grey-scale}),
velocity ({\it arrows}), temperature
of amorphous carbon grains ({\it solid black contour lines}), and
temperature of silicate grains ({\it dotted contour lines}) for the
60\,M$_\odot$ case of Yorke \& Sonnhalter (2002) at $t=25,000$~yr,
after which 33~M$_\odot$ have accreted onto the core. 
{\bf Right:} Time dependent mass accretion rate of central object for
60~M$_\odot$ case shown at left.
}
\end{figure}

\begin{figure}
\plotone{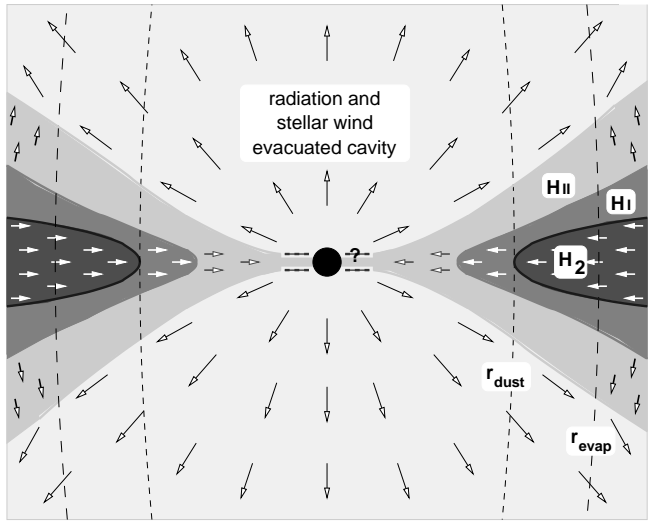}
\caption{The inner accretion disk: inward radial flow is 
allowed in the equatorial plane.  A polar cavity is evacuated by a
combination of radiation and the stellar
wind. The disk itself is self-shielded from the intense EUV field
by an ionization front separating H{\sc ii} and H{\sc i} gas
and from the FUV field interior to the H{\sc i}/H$_2$
interface by dust, molecular hydrogen, and CO.  The dust is destroyed
at $r_{\rm dust}$. Interior to  $r_{\rm evap}$ even the ionized gas
is gravitationally bound.
}
\end{figure}

One can speculate on the affect that outflows will have on the
accretion through an accretion disk.
The inner part of the accretion disk could well look like the
configuration shown in Fig.~9.  Radiation and the stellar wind
from the central star (presumably already hydrogen-burning) 
evacuate a cavity in the polar direction as shown in the example
depicted in Fig.~8. At the interface
between the supersonic outflowing stellar wind and the denser
subsonic H{\sc ii} disk atmosphere some disk material will be
removed but it is unlikely that this can prevent inward flow of disk
material.  Inward radial flow of dusty molecular gas is allowed
in the equatorial plane of the disk (assuming
angular momentum is transfered outward) because of the low
radially directed radiation flux (flashlight effect).  Angular
momentum transfer could result from either magnetic fields in
the disk or from the tidal effect of nearby stars.  Indeed, rapid
accretion through a disk may be a direct consequence of having
nearby companions, thus explaining why massive stars are generally
members of multiple systems.

Once the disk material crosses
$r_{\rm dust} \sim 25\,{\rm AU}\,[M_*/30\,{\rm M}_\odot]^{1.6}$,
the radius of dust
destruction, its opacity decreases and it is not easily stopped
by radiation.  It is, however, still unclear how the
disk material ultimately flows onto the star.  Surely, the
disk ``puffs up'' close to the star, in analogy to the accretion
disks in AGNs.
Outside of $r_{\rm evap} \sim 130\,{\rm AU}\,[M_*/30\,{\rm M}_\odot]$,
where the escape velocity is less than 10~km~s$^{-1}$, 
the disk loses material via photoevaporation on a time scale
of $\sim$10$^5$~yr. This is of the same order as the
accretion time scale; these competing effects will determine
the final mass of the star.

\section{Discussion and Conclusions}

Massive stars can in principle be formed via accretion through a disk.
An accreting star quickly evolves to the main sequence after about
10~M$_\odot$ have accreted, but the star is not readily observable
as such.  Still obscured by the material in the vicinity, the
appearance will be that of an UCH{\sc ii}.  Radiative acceleration,
photoevaporation, and stellar winds eventually destroy the disks,
but prior to this accretion onto
the star provides an additional source of luminosity.  This accretion
is expected to be highly variable and episodic.
A powerful radiation-driven outflow in the polar
directions and a ``puffed-up'' (thick) disk result from the high
luminosity of the central source.  The details of how disk material
ultimately flows onto the star is still unclear --- as is often
the case for accretion disks in general.
 
In this report I have not addressed the issues of embryo multiple
systems (binaries, etc.) which compete for accretable material.
Future studies will have to address this issue as well as the
interactions of nearby disks with multiple massive
stars to provide a more realistic picture of massive
star formation.

Accretion via accretion disks is a universal astrophysical phenomenon,
which operates at scales ranging from the central machines of AGNs to
X-ray binaries and perhaps to planets.  Why should accreting massive
stars be an exception?

\begin{acknowledgments}
This research was conducted at the Jet Propulsion Laboratory,
California Institute of Technology and has been supported by the National
Aeronautics and Space Administration (NASA) under the auspices of
the ``Origins'' Program and grant NRA-99-01-ATP-065.
\end{acknowledgments}


\begin{references}
\reference Behrend, R., Maeder, A. 2001, A\&A, 373, 190
\reference Bonnell, I.A., Bate, M.R., Zinnecker, H. 1998 MNRAS, 298, 93
\reference Bonnell, I.A., 2002, PASP, 267, ed. P.A. Crowther, 193
\reference Cabrit,  S., Bertout, C. 1992, A\&A, 261, 274
\reference D'Antona, F., \& Mazzitelli, I. 1994, ApJS, 90, 467
\reference Eiroa, C., Casali, M. M., Miranda, L. F., Ortiz, E. 1994,
 A\&A, 290, 599
\reference Hollenbach, D., Yorke, H.W., Johnstone, D. 2000, in
 {\sl Protostars and Planets IV},
 eds. Mannings, Boss, Russell, (Tucson: Univ. Ariz. Press), p. 401
\reference Iben, I., Jr. 1965, ApJ, 141, 993
\reference Kahn F.D. 1974, A\&A, 37, 149
\reference Maeder, A, 2002, In: Highlights of Astronomy, Vol. 12,
ed. H. Rickman, p. 170
\reference Mathis, J.S., Rumpl, W., Nordsieck, K.H. 1977, ApJ, 217, 425
\reference Martin-Pintado, J., Neri, R., Thum, C., Planesas, P., Bachiller, R.
 1994, A\&A, 286, 890
\reference McKee, C.F., Tan, J.C., 2003, ApJ, 585, 850
\reference Meynet, G., Maeder, A. 2000, A\&A, 361, 101
\reference Nakano T., Hasegawa T., Norman C. 1995, ApJ 450, 183
\reference Preibisch, T., Ossenkopf, V., Yorke, H.W., Henning, T., 1993,
 A\&A, 279, 577
\reference Richer, J. S., Shepherd, D. S., Cabrit, S., Bachiller, R.,
 Churchwell, E. 2000,
 in {\sl Protostars \& Planets IV}, ob. cit., p. 867
\reference Richling, S., Yorke, H.W. 1997, A\&A, 327, 317
\reference Shepherd, D.S., Churchwell, E. 1996, \apj, 472, 225
\reference Shepherd, D.S., Claussen, M.J., Kurtz, S.E., 2002,
PASP, 267, ed. P.A. Crowther, 415
\reference Suttner G., Yorke H.W. 2001, ApJ, 551, 461
\reference Wolfire M.G., Cassinelli J.P. 1987, ApJ, 319, 850
\reference Yorke, H.W., 1986, ARAA, 24, 48
\reference Yorke, H.W., 2002, PASP, 267, ed. P.A. Crowther, 165
\reference Yorke, H.W., Bodenheimer, P. 1999, ApJ, 525, 330 
\reference Yorke, H.W., Kr\"ugel, E. 1977, A\&A, 54, 183
\reference Yorke, H.W., Sonnhalter, C., 2002, ApJ, 569, 846
\end{references}
\end{document}